\documentclass[conference]{IEEEtran}

\usepackage[dvips]{graphicx}
\usepackage{amsmath,amssymb}
\usepackage{algorithm}
\usepackage{algorithmic}
\usepackage{xcolor}
\usepackage{hyperref}
\hypersetup{
  colorlinks   = true, 
  urlcolor     = black, 
  linkcolor    = black, 
  citecolor   = black 
}
\usepackage{url}
\usepackage[sort&compress,numbers]{natbib}

\begin{document}
\title{Facilitating the Manual Annotation of Sounds When Using Large Taxonomies}
\date{}

\author{
\IEEEauthorblockN{Xavier Favory, Eduardo Fonseca, Frederic Font, Xavier Serra}
\IEEEauthorblockA{Music Technology Group - Universitat Pompeu Fabra \\ Barcelona, Spain \\ {name.surname}@upf.edu\\}
}
\maketitle

\begin{abstract}

Properly annotated multimedia content is crucial for supporting advances in many Information Retrieval applications.
It enables, for instance, the development of automatic tools for the annotation of large and diverse multimedia collections.
In the context of everyday sounds and online collections, the content to describe is very diverse and involves many different types of concepts, often organised in large hierarchical structures called taxonomies.
This makes the task of manually annotating content arduous.
In this paper, we present our user-centered development of two tools for the manual annotation of audio content from a wide range of types.
We conducted a preliminary evaluation of functional prototypes involving real users. The goal is to evaluate them in a real context, engage in discussions with users, and inspire new ideas.
A qualitative analysis was carried out including usability questionnaires and semi-structured interviews. This revealed interesting aspects to consider when developing tools for the manual annotation of audio content with labels drawn from large hierarchical taxonomies.

\end{abstract}

\section{Introduction}
Accessing multimedia content is one of the core challenges in multimedia research. 
In the past decades, automatic content description methods have proliferated and can be used, with different accuracies, for detecting semantic concepts from low-level features derived from the content digital representation.
However, there is a persistent \textit{semantic gap}~\cite{celma2006bridging} produced by the lack of accordance between the information that can be extracted from the data and the interpretation that the same data has for a user.

Nowadays, successful automatic description methods are based on approaches that often rely on a lot of data for training and evaluation.
As a consequence, manual generation of content description is of high importance for the realisation of intelligent systems able to produce meaningful automatic content descriptions.
%
Recent advancements partially come from the popularity of online sharing platforms, which made available a large amount of data~\cite{russakovsky2015imagenet}.
%
%
%
In these platforms, description and tagging systems have become increasingly popular.
Users can add textual descriptions or keywords (i.e., tags) to Internet resources (e.g., web pages, images, music) without relying on a controlled vocabulary. 
This makes it less demanding for users than, for example, classifying objects into predefined categories.
Although these user-generated descriptions enable the development of valuable searching tools for online-shared content \cite{marlow2006ht06}, they are not always directly adequate for the effective management of multimedia content.
Indeed, the interoperability of the content descriptions is fundamental to information sharing, exchange and reuse. 
Therefore, having semantic content metadata that is understandable and processable both by machines and humans is crucial.
%

To address this issue, taxonomies allow to organise and structure concepts.
In the audio-related fields they are the first step towards the classification of sounds into groups based on different subjective or contextual properties \cite{schafer1993soundscape}.
Disparate taxonomies have been developed based on subjective similarity, sound source or common environmental context.
%
%
%
%
%
However, since sounds are multimodal, multicultural and multifaceted, there is not a common taxonomy that allows to organise large and diverse sound collections.
%
%
%
%
%
%
%
%
%
%
%
%
%
%
Some works proposed taxonomies for environmental sounds, based on the interaction of materials \cite{gaver1993world} or according to their physical characteristics \cite{schafer1993soundscape}.
More recent research on studying soundscapes shows that the taxonomical categorisation of environmental sounds is not trivial and involves many different fields, e.g., human perception or urban design \cite{brown2011towards, salamon2014dataset}.
%
For musical content, many music genre taxonomies appeared from the Music Industry and its consumers. Yet no standard taxonomy has been established since it depends highly on our cultural contexts. In fact, each distributor has his own strategy towards its targeted market \cite{pachet2000taxonomy}.    

Despite all the accomplishment in designing specific taxonomies, 
the creation of larger, general-purpose taxonomies has recently gained attention among the research community~\cite{gemmeke2017audio}.
Instead of focusing on the recognition of a specific subset of sounds, general-purpose taxonomies enable tasks that aim to recognise and describe a wider (and usually more generic) range of sounds \cite{fonseca2018general}.
Methods to solve these tasks are desirable, for example, in environments such as smart buildings or smart cities and more generally in IoT applications.
%
%
Another application is the automatic description of multimedia content in the context of large online collections like Freesound (\url{https://freesound.org/}) \cite{font2013freesound} or Youtube. 
This can enable the enhanced organisation and retrieval of multimedia content, thus making it more accessible to the public.
In these cases, training general-purpose systems with large-vocabulary audio datasets seems more suitable to be able to describe a wide variety of content types.

The recently released AudioSet Ontology proposes one of the biggest taxonomies which structures 632 audio-related categories~\cite{gemmeke2017audio}.
Rather than being domain-specific, it contains the most common concepts used for describing everyday sounds.
AudioSet has a companion website that includes a web-interface to navigate through the taxonomy and listen to sound examples, which provides an overview of its content (\url{https://research.google.com/audioset/}).
Sounds are related to many things, such as nature, urban design, music and culture. 
Consequently, sound related taxonomies are supposed to evolve and adapt, and it is important for people to understand, use and discuss about them.
For this reason, proposing tools and interfaces for browsing taxonomies would lead to vast advancements in the many related fields.
Likewise, these tools can assist the annotation of the content in online sharing platforms, which would facilitate its use for research or multimedia sharing.

In this paper, we advance our user-centered design process of proposing general-purpose annotation tools that can be used for annotating all sorts of audio content.
We take advantage of the AudioSet Ontology which provides a hierarchical taxonomy of very broad acoustic categories.
The main goal is to facilitate the exploration and use of predefined categories taken from large taxonomies.
%
In section II, we first explain the context of this work by briefly presenting the current outcome of the Freesound Datasets initiative \cite{fonseca2017freesound}, which focuses on the annotation of sounds from the Freesound database.
We then motivate the need of two tools for the manual annotation of audio samples.
In section III, we describe the two annotation tools we developed: one allows to add labels to audio samples, and another allows to refine previously existing labels.
In section IV, we present a preliminary evaluation of the two tools carried out with real users.
We conclude the paper in section V.

\section{Motivations}
\subsection{Context}

In previous work \cite{fonseca2017freesound}, the authors describe FSD, a large-scale open audio dataset based on Freesound content annotated with categories drawn from the AudioSet Ontology.
%
%
Currently, FSD presents annotations that express the presence of a sound category in audio samples.
The creation of FSD started with the automatic population of each category in the AudioSet Ontology with a number of candidate audio samples from Freesound. 
This process automatically generated over 600k candidate annotations.

\begin{figure}[h]
	\centering
	\includegraphics[width=0.50\textwidth]{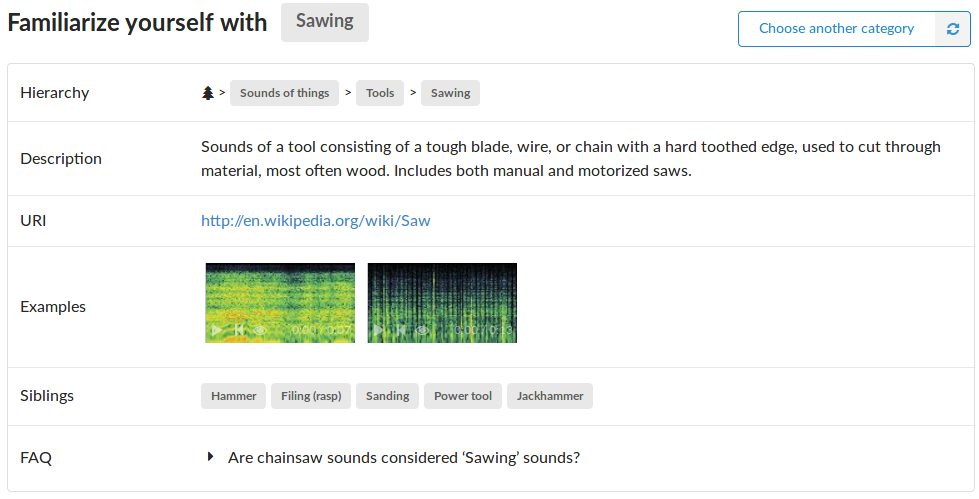}
	\caption{Screenshot of the familiarisation interface of the Freesound Datasets platform validation task}
	\label{f}
\end{figure}

To verify the validity of these automatically generated annotations, we developed a validation tool with an interface that helps users to understand a category and its context in the AudioSet Ontology.
This validation tool is deployed in the Freesound Datasets platform (\url{https://datasets.freesound.org/}).  Fig. \ref{f}  shows the part of the interface used for the familiarisation of a user with a category. It displays information such as the name, description, sibling and children categories of a specific category.

\subsection{Motivating new annotation interfaces}

This approach produced a considerable amount of annotations which already helped communities of researchers to investigate new machine learning methods \cite{fonseca2018general}.
However, generating annotations automatically presents a number of shortcomings. 
For instance, an automatic process can generate incorrect or not specific labels, and it can also fail to generate some labels.
We argue that the usefulness and reliability of datasets increase with the proximity of its annotations towards what we denote as \textit{complete} or \textit{exhaustive} labeling (i.e., all the acoustic material present in the audio file is annotated).

To achieve this complete labeling status through manual annotation, a number of actions would be required. 
First, assuming the existence of automatically generated annotations, it would be needed to validate them. 
Then, missing labels should be \textit{generated}, and generic or unspecific labels should be further \textit{refined}.
The two annotation tools presented in the next section address the two latter issues.

\section{The annotation tools}

\begin{figure}[h]
	\centering
	\includegraphics[width=0.50\textwidth]{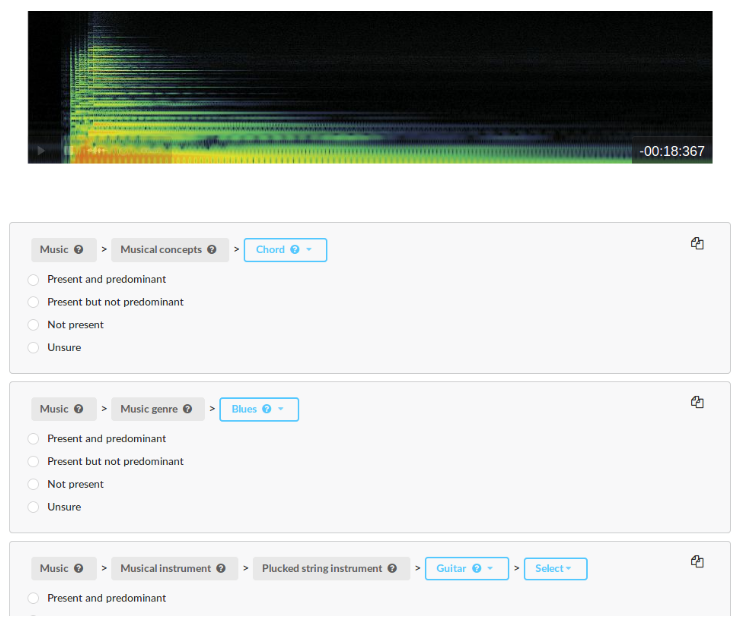}
	\caption{Screenshot of the Audio Commons Refinement Annotator displaying a sound sample and its three suggested label paths}
	\label{f4}
\end{figure}

In this section we describe the two novel interfaces that we developed. The code is available at: \url{https://github.com/MTG/freesound-datasets/tree/annotation-tools-FRUCT2018/}. Both tools are implemented mostly with web client languages, which allows their easy integration in other projects. 
The Audio Commons Manual Annotator (AC Manual Annotator) aims at adding missing labels, whereas the Audio Commons Refinement Annotator (AC Refinement Annotator) allows to refine and specify existing labels.
These tools can be useful not only to annotate during a post-precessing stage, like in Freesound Datasets, but also to provide annotations when a user publishes content in an online platform such as Freesound.
Both of the tools focus on annotating a single sound resource at a time.
The audio content is accessible from a player displaying the spectrogram of the sound, which can facilitate the localisation and recognition of sound events in the clip (Fig. \ref{f1} \& \ref{f4}) \cite{cartwright2017seeing}. 

\subsection{Generate annotations}

\begin{figure}[h]
	\centering
	\includegraphics[width=0.50\textwidth]{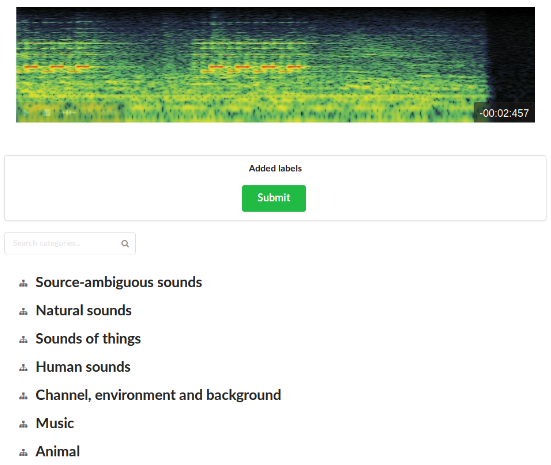}
	\caption{Screenshot of the Audio Commons Manual Annotator}
	\label{f1}
\end{figure}

\begin{figure}[h]
	\centering
	\includegraphics[width=0.50\textwidth]{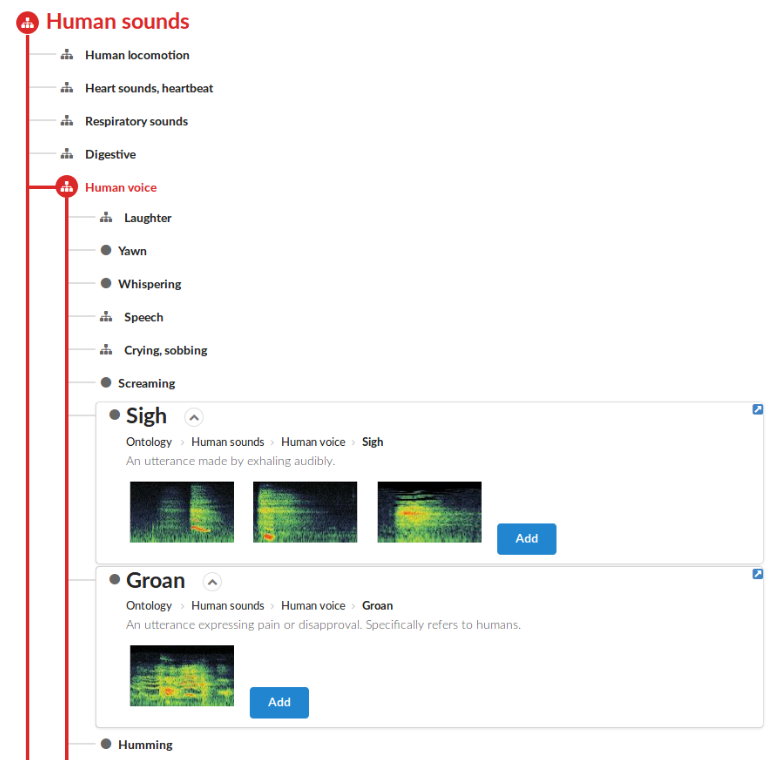}
	\caption{Screenshot of the Audio Commons Manual Annotator taxonomy table, showing the descriptions and examples of “Sigh” and “Groan”, together with their hierarchy location}
	\label{f3}
\end{figure}

With the AC Manual Annotator, labels can be assigned to an audio clip. 
The main idea behind this interface is to provide a way to facilitate the quick overview of categories.
Moreover, considering the large size of the hierarchical structure in taxonomies like AudioSet, it is important to show the location and context of the categories within the hierarchy. 
Another design criteria was to allow the comparison of different categories by simultaneously displaying their information.
In the proposed interface, a text-based search allows to locate categories in the taxonomy table.
We used text from the category names and descriptions to perform some trigram based queries (a feature that Postgres, our database backend, implements).
The taxonomy table allows users to open parts of the taxonomy in order to visualise children categories simultaneously. 
For each category, textual descriptions are shown, along with sound examples when available (Fig. \ref{f3}).

\pagebreak

A typical use workflow would consist in:
%
%
\vspace{-0.3em}
\begin{itemize}
\setlength\itemsep{0.2em}
\item Listen to the sound sample (Fig. \ref{f1}, top).
\item Use the text-based search to locate categories in the taxonomy table (Fig. \ref{f1}).
\item Explore the taxonomy table to understand well the located category, and perhaps find other more relevant categories (Fig. \ref{f3}).
\end{itemize}

\begin{figure}[h]
	\centering
	\includegraphics[width=0.50\textwidth]{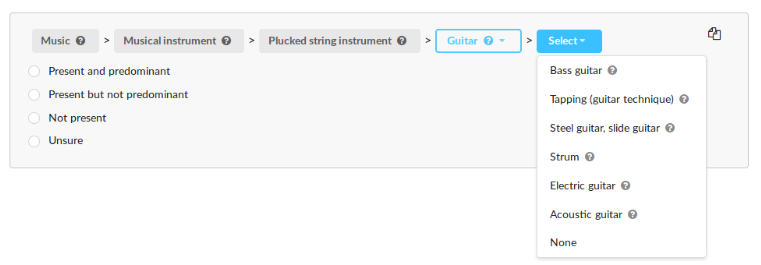}
	\caption{Screenshot of the Audio Commons Refinement Annotator showing a dropdown displaying the children categories of “Guitar”}
	\label{f5}
\end{figure}

\begin{figure}[h]
	\centering
	\includegraphics[width=0.50\textwidth]{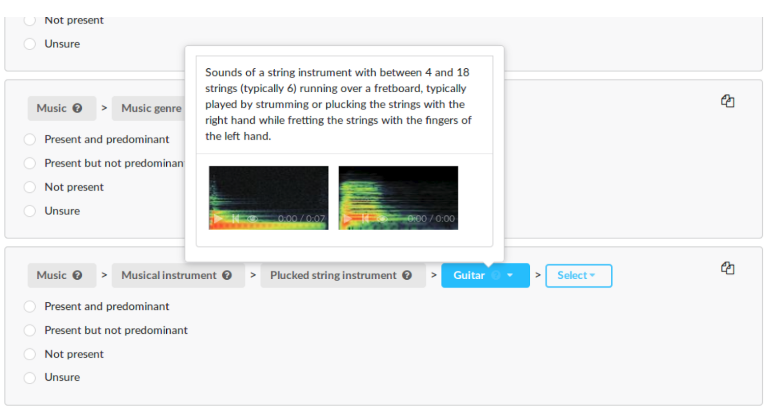}
	\caption{Screenshot of the Audio Commons Refinement Annotator showing the description and examples of the “Guitar” category in a popup}
	\label{f6}
\end{figure}

\subsection{Refine annotations}

The AC Refinement Annotator displays some previously existing labels as rows, as it can be seen in Fig. \ref{f4}. 
The annotator can examine their location in the AudioSet hierarchy as well as their siblings and children categories. 
By making use of the hierarchy, the main goal of this tool is to aid the annotation process by providing an iterative way of specifying the type or nature of the content.
Fig. \ref{f5} shows how the children categories of the proposed label ``Guitar" are displayed in a dropdown, which allows to modify the label and define it more precisely. 
For every label, popups show the category description and examples when available (Fig. \ref{f6}). 
Moreover, it is possible to duplicate a label using the icon at the top right corner of a label path. 
This allows, for instance, to specify a label by adding two of its children categories.
In the final step of the refinement process, the user is asked to verify the \textit{presenceness} of the selected category in the audio clip. 

A typical use workflow would consist in:
\vspace{-0.3em}
\begin{itemize}
\setlength\itemsep{0.2em}
\item Listen to the sound sample (Fig. \ref{f4}, top).
\item Inspect the proposed labels (Fig. \ref{f4}).
\item Refine the proposed labels by inspecting the related siblings and children (Fig. \ref{f5} \& \ref{f6}).
\item Validate the presence of the proposed or refined category.
\end{itemize}

\section{Preliminary evaluation}

In the context of sound collections annotation, there is a need for proposing new manual interfaces to properly annotate audio content, with labels that are comparable and of the same nature. 
In this experiment, we present our user-centered design process on the development of novel tools for annotating audio content from a wide variety of types. 
We use the annotator tools as \textit{technology probes} to observe their use in a real context, to evaluate their functionalities and to inspire new ideas \cite{hutchinson2003technology}. 

\subsection{Methodology}
We gathered eight participants with different levels of expertise.
Each one of them was provided with one of the tools and was asked to annotate a list of sounds one by one.
%
%
We selected sounds from the Freesound Datasets platform featuring one or more of the following aspects: (i) containing multiple sound sources, (ii) presenting background noise or (iii) being hard to recognise. 
This process resulted in a list of 9 and 15 sounds for the generation and refinement tools respectively.
Some guidelines were shown to them, together with verbal explanations given by the examinator.
%
%
At the end of the task, they were provided with a questionnaire containing some usability and engagement questions.
Finally, semi-structured interviews were carried out, including open-ended questions as well as specific questions related to observed behaviors during the development of the task.
This enables discussion using thematic analysis in order to identify emerging themes from participants' answers.

\subsection{Results and discussion}

\textbf{Finding a category in the taxonomy.} 
It is essential to provide ways for efficiently browsing and exploring such an extensive set of audio categories. 
Text-based search provides a way for people to find categories with their own words. 
This is particularly efficient when the annotator recognises the sources and want to quickly add the corresponding audio category to the content. 
As a way to improve the retrieval from the text-based search, one participant proposed to add some of the children of the retrieved categories to the results.
This option was tested when developing the search engine, but was discarded because it tended to add a lot of results which made the localisation of the relevant categories harder. 
Moreover, we could also use external lexical resources such as WordNet or Wikipedia to improve system's recall, by using synonyms terms and page content terms respectively.

However, text-based search can fail when the annotator is not familiar with the vocabulary. 
%
%
She can then rely on the hierarchical structure of the categories. 
Tree visualisations are a direct representation of it, and can help by allowing to iteratively define more precise concepts starting from the broader upper levels of the taxonomy.
As well, tables are a natural way for browsing collections of items. 
The taxonomy table we provided in the AC Manual Annotator aims at combining tree and table structures in order to allow efficient and fast exploration of the categories.
Moreover, locating similar categories close from each other helps to refine and validate the choice of a category (especially for categories that are almost identical and differ only in small details).

\textbf{Exploring the taxonomy.} 
The hierarchy structuring the audio related concepts assumes that deep located categories convey more information that the others. 
%
%
Therefore, it is important to use labels as specific as possible in order to accurately describe the audio content. 
When using the AC Refinement Annotator, some participants showed interest in seeing all the hierarchy at once. 
However, we believe that the task is facilitated if only the relevant context for each step of the iterative process is shown.
Specifying labels in an iterative fashion (i.e., progressively, such that their meaning is narrowed down in every step) seems to be helpful. 
It can ease and speed up the generation of accurate labels by focusing on the most relevant semantic audio aspects.
Nonetheless, during the navigation through the different levels of specificity in the hierarchy, a participant was sometimes not inspecting, or hesitating to check, the children of a category. 
This occurred due to several reasons: (i) since no sound examples were available in the present category, he assumed this would also be the case in deeper hierarchy levels.
Hence, he decided not to explore this branch due to lack of confidence with it; (ii) he also assumed that since  the original category was not appropriate, none of the children would be either (where in fact, one of them was). 
The AC Manual Annotator mitigates this problem and facilitates quick inspection of the categories, since the children can be automatically displayed when a category is selected in the taxonomy tree.

\textbf{Difficulty in recognising a sound identity.}
In the context of post-processing annotations of audio content, the annotator is typically not the publisher of the content.  
Hence, the annotator usually does not know the details of the recording conditions or what sound sources were captured.  
Furthermore, listening to the sound does not necessarily lead to the identification of the sound source(s) as it can sometimes be a very complex task.
Under these circumstances, for the audio content that annotators were not able to recognize, the following behaviors were observed.
When using the AC Manual Annotator, the annotators tended to choose abstract categories that do not convey the source identity, but rather some other aspects of the sound source  (e.g., onomatopoeic labels that phonetically imitate, resemble, or suggest the sound it describes).
In the AC Refinement Annotator tool, where participants were guided towards the identification and specification of the sources, they usually stopped at a certain hierarchical level, thus providing some imprecise labels.
%
%
As expected, labels gathered with the \textit{generation} tool were much more different than those gathered with the \textit{refinement} tool.
One of the reason was that with the AC Manual Annotator tool, users chose different abstract labels for describing the content, since their exact meaning seems to vary across annotators.
%
%
%
%
%
%
%
%

To improve the consistency of the produced labels, it was discussed to give access to the metadata that often accompany online shared media, e.g., title, description and tags. 
These informations can guide annotators on understanding the context and providing more accurate annotations.
However, some participants argued that these informations should not be given at first. 
For them, access to metadata should be an additional aid that could be requested only after having spent a certain effort on analysing the audio content.
Providing directly the metadata would correspond more to a transcription task, where annotators could focus only on the metadata, and forget some important sound aspects that the metadata fail to convey.
%

\textbf{The annotators' commitment is highly variable.} 
In addition to the precision of labels, the AC Refinement Annotator also allows to explore siblings categories that can sometimes correspond to slightly different concepts.
This enables correcting the, potentially noisy, automatically generated labels.
However, this feature led to variable results in terms of labels produced and time spent annotating.
Users of the \textit{refinement} annotator spent from 35 minutes to 1h20 annotating 15 sounds.
Some participants put a lot of efforts exploring sibling categories in the hierarchy, making them waste time when considering the amount of refined labels (from 23 to 34 labels with a present validation).
In contrast, the users of the AC Manual Annotator spent from 25 to 30 minutes performing the task.
%

Finally, it was observed that some participants gave a lot of importance to category sound examples and children, rather than relying on the name and textual description.
This presents a risk since, in many occasions, neither the sound examples nor the listed children can be fully representative of a category diversity and complexity.
It is therefore important that the tools promote the utilization of all the available information for annotators to take more solid and reliable decisions.

\section{Conclusions}
In this paper we motivated the need for novel interfaces that facilitate the use of categories from large-scale taxonomies when annotating audio content.
We presented the context of the Freesound Datasets initiative, which aims at creating openly available audio datasets.
Two annotation interfaces were presented, which allow to target specific shortcomings when automatically generating labels.
A preliminary evaluation with users allowed to evaluate our first versions of the tools and engage discussions.
%
%
%

Future work should focus on making the tasks faster, and aid the annotators on producing more exhaustive and consistent annotations.
It will include improvements on the design, such as making the sound player more reachable to allow simultaneous exploration of category examples and comparison with the audio resource being annotated.
%
%
Simplification of the AC Refinement Annotator task by disabling the exploration of sibling categories in the taxonomy hierarchy.
%
In addition, improved and more detailed task instructions should be designed, containing specific indications to make users focus on specific sound aspects covered by the taxonomy.  These measures could help annotators to produce more comprehensive annotations.

{\small
\section*{Acknowledgment}
This work has received funding from the European Union's Horizon 2020 research and innovation programme under grant agreement No 688382 ``AudioCommons'' and from a Google Faculty Research Award 2017. The authors thank Lorenzo Romanelli for his help with the development of the annotation tools, and the participants of the evaluation for the valuable feedback gathered.
}

{\small
\bibliography{refs}}
\bibliographystyle{unsrtnat}

\end{document}